# Magnetic Effects on Dielectric and Polarization Behavior of Multiferroic Hetrostructures

Sandra Dussan,[1] Ashok Kumar,[1] J. F. Scott,[1,2] and Ram S. Katiyar[1*]

[1]Department of Physics and Institute for Functional Nanomaterials, University of Puerto Rico, San Juan, Puerto Rico 00931-3343, USA.

[2]Cavendish Laboratory, Dept. Physics, Cambridge University, Cambridge CB3 0HE, U. K.

**ABSTRACT**

$PbZr_{0.52}Ti_{0.48}O_3/La_{0.67}Sr_{0.33}MnO_3$(PZT/LSMO) bilayer with surface roughness ~ 1.8 nm thin films have been grown by pulsed laser deposition on $LaAlO_3$(LAO) substrates. High remnant polarization (30-54 $\mu C/cm^2$), dielectric constant(400-1700), and well saturated magnetization were observed depending upon the deposition temperature of the ferromagnetic layer and applied frequencies. Giant frequency-dependent change in dielectric constant and loss were observed above the ferromagnetic-paramagnetic temperature. The frequency dependent dielectric anomalies are attributed to the change in metallic and magnetic nature of LSMO and also the interfacial effect across the bilayer; an enhanced magnetoelectric interaction may be due to the Parish-Littlewood mechanism of inhomogeneity near the metal-dielectric interface.

**Corresponding Author: E-mail :(*)** *rkatiyar@uprrp.edu*

Multiferroic materials were defined as a class of functional materials that combine two or more functional properties -- ferromagnetism, ferroelectricity and ferroelasticity;[1,2] However, more recently the term has been used with a narrower meaning describing only those materials which combine magnetic and ferroelectric ordering (since most ferroelectrics care automatically ferroelastic by symmetry). The simultaneous existence of ferroelectric and ferromagnetic properties in the same material have attracted scientific and technological interest in the last decade due potential applications in magnetic data storage, sensors, non-volatile memories, actuators, micromechanical applications, and the emerging field of spintronics.[3,4] Many ferromagnetic-ferroelastic (magnetostrictive) and ferroelectric-ferroelastic (piezoelectric) materials are known, but very few multiferroic ferroelectric-ferromagnetic single-phase material exist in nature, which derives from the fact that the two properties are often chemically incompatible; e.g., in oxide perovskites the existence of transition metal d-electrons are essential for ferromagnetism, but work against the tendency for off-center displacement of cations that cause ferroelectricity.[5] Alternative mechanisms to obtain both electrical and magnetic ordering in a single-phase material have included the construction of artificial multiferroic thin films sandwiches combining a ferroelectric and a ferromagnetic material.[6] Layers of the perovskite $PbZr_xTi_{1-x}O_3$ (PZT) as a constituent of heterostructures have often been used, owing to its high remanent polarization ($P_r$), low coercive field ($E_c$), and high Curie temperature ($T_c$).[7] To complete the sandwich, $La_xSr_{1-x}MnO_3$ is in an interesting candidate not only as a ferromagnetic-layer element in the heterostructures but also as the bottom electrode for ferroelectric films, due to high Curie temperature ~340 K and its lattice parameter (~ 3.87 Å) which closely matches that of the perovskite ferroelectric material. Additionally, LSMO is a half-metallic material, and it exhibits colossal magnetoresistance (CMR).[8,9] Previous work reported good magnetic and ferroelectric properties for $PbZr_xTi_{1-x}O_3/La_{0.7}Sr_{0.3}MnO_3$ heterostructures deposited on $SrTiO_3$ (100) substrates.[10] The magnetoelectric effect (ME) has been observed in epitaxial $PbZr_{0.52}Ti_{0.48}O_3/La_{0.7}Sr_{0.3}MnO_3$ and $PbZr_{0.3}Ti_{0.7}O_3$ on $La_{1.2}Sr_{1.8}Mn_2O_7$ single-crystal substrates.[11,12] In the present work we describe the magnetic effect on electrical properties of epitaxial bi-layered heterostructure. The frequency and temperature dependent dielectric anomalies will be described in the light of the ferromagnetic-paramagnetic transition in LSMO.

Stoichiometric ceramic targets of $La_{0.67}Sr_{0.33}MnO_3$ (LSMO) and $PbZr_{0.52}Ti_{0.48}O_3$ (PZT) with 20% excess of lead oxide (to compensate for loss of volatile Pb) were synthesized by a conventional solid-state route. Three different PZT/LSMO bilayered films with varying deposition temperature of LSMO were fabricated by pulsed laser deposition (PLD) employing a KrF excimer laser ($\lambda$=240 nm). The LSMO layer was grown at $700^0C$ and $600^0C$ under an oxygen pressure of 80 mTorr, using a laser energy density of (1.8 J/cm$^2$) and repetition rate of 10 Hz, followed by annealing at $700^0C$ for 30 minutes in oxygen. The PZT layer was then deposited on the LSMO layer at $600^0C$ at the same conditions. After deposition the heterostructure was annealed at $700^0C$ for 30 minutes in oxygen at a pressure of 300Torr. Finally, the films were cooled down to room temperature at a slow rate. The total thicknesses of the films were around 550 nm. The structural analysis was done with a Siemens D500 x-ray diffractometer (Cu K$\alpha$ radiation) in a $\theta$-$2\theta$ scan. The atomic force microscopy (Veeco-AFM-contact mode) was used to examine the morphology and the surface roughness. The film thickness was determined using an X-P-200 profilometer. The electrical measurements were carried out on capacitor structures fabricated with Pt top electrode (diameter of ~200$\mu$m) deposited by DC sputtering through a shadow mask. The capacitance and loss spectra were measured with an impedance analyzer (HP4294A) from 100Hz to 1MHz. Ferroelectric hysteresis loops were measured utilizing a RT6000HVS probe unit (Radiant) in the virtual ground mode. The magnetic measurements were performed using a vibrating-sample magnetometer (VSM-lakeshore 736) at room temperature.

Figure 1 shows the x-ray diffraction patterns of PZT (550 nm)/LSMO (200 nm) heterostructures deposited at two different temperature. For the first specimen the deposition temperature of LSMO was $700^0$C and that of the PZT was $600^0$C. We denote in this paper heterostructures with LSMO annealed at $700^0$C as HS1 and those annealed at $600^0$C as HS2; but in all cases the PZT was annealed at $600^0$C. It may be observed in Fig.1 that the all peaks corresponding to pure PZT and LSMO are present. In both cases only the ($\ell$00) diffraction peaks of LSMO and PZT were observed, along with those of the single crystal $LaAlO_3$ substrate. This suggests that the individual ferromagnetic and ferroelectric layers were grown epitaxially block-by-block on the substrate and formed smooth films without any secondary parasitic phases. The AFM images of the PZT/LSMO heterostructures are shown as an inset in Figure 1. The surface morphology of the films did not present any evidence of cracking or defects (pitting or pores), indicating fairly uniform deposition. The topographic view clearly shows spherical grains in epitaxially grown films on $LaAlO_3$ substrates. Film smoothness was verified by their low root-mean-square (rms) surface roughness values as 1.8 nm.

The frequency dispersion of dielectric constant ($\varepsilon$) and loss tangent (tan $\delta$) of PZT/LSM heterostructures (HS) as a function of temperatures and frequencies are shown in the Figure 2a (heterostructured HS1), 2b (heterostructure HS2). They reveal strong frequency dispersion with a drastic decrease in the $\varepsilon$ value at frequencies above $10^5$ Hz for HS1 and HS2. There are two major factors influencing the dielectric dispersion. The first one is related to extrinsic sources. These extrinsic sources can be attributed to the additional capacitance arising from grain boundary and interfaces between layers as a consequence of the contribution of two different materials (ferroelectric and ferromagnetic). In all cases the dielectric constant increases with the rise in temperature in the frequency range from $10^2$ Hz to $10^5$ Hz. The loss tangents were nearly constant (~0.04) at low and room temperature in the frequency range $10^2$ Hz -$10^4$ Hz and increasing to as much as ~3.0 for high temperatures (600 K) at frequencies above $10^5$ Hz. These frequencies correspond to the drastic change in ($\varepsilon$) for the heterostructures (These high frequency increases cannot arise from inductances due to lead wires, because they are absent in nonmagnetic materials tested such as pure PZT; nor do they decrease if the leads are shortened). The second potential cause is related to the ferromagnetic-metal to paramagnetic-insulator (M-I) transition of the LSMO (which also forms the bottom electrode). It is interesting to note that we observed huge frequency dispersion at temperatures above the metal-insulator transition. The high-frequency dielectric constant is very low whereas at low frequencies a very high dielectric constant was observed. This indicates that LSMO still behaves as metallic at sufficiently low frequencies at temperatures slightly above its M-I transition, whereas it acts insulating at higher frequencies (>$10^4$-$10^5$ Hz). A sharp dielectric anomaly occurs at the M-I transition. The dielectric constant drops abruptly at higher frequencies, whereas it behaves normally at lower frequencies (<10 KHz). Dielectric tangents show a high loss for high frequency and low loss for lower frequencies. These observations are explained as follows: (i) LSMO thin films exhibit a rather diffuse M-I transition over a wide range of temperature; (ii) the M-I transition is much sharper in films with higher annealing temperatures, and this leads to a low dielectric constant at higher frequency (40 @ 1 MHz); (iii) both magnetization and polarization were observed to be smaller at low annealing temperatures of LSMO; (iv) larger dielectric constants at higher frequencies are seen in these heterostructures and suggest that high conductivity of LSMO occurs with low annealing temperatures; (v) Remanent polarization $P_r$ is enhanced monotonically with decrease in temperature (below the M-I temperature), which clearly indicates improvement in metallic behavior of LSMO (this in turn increases the polarization). Finally, giant magneto-capacitive responses in PZT/LSMO hetrostructure in the CMR regions suggests that if it is intrinsic effects due to temperature dependent dynamic magneto-capacitive coupling, it would be of great interest as a magnetic field sensor, otherwise, extrinsic effects due to inhomogenity near the interface.

A more comprehensive analysis of the effects of the metal-dielectric interface on the magnetic and dielectric properties of our heterojunctions might be afforded by the theory of Parish and Littlewood.[13] These authors show that interdiffusion of atoms at a metal-insulator interface to produce an inhomogeneous medium can cause strong enhancements of the magnetic interaction with polarization via strain.

Figure 3 shows the room-temperature ferroelectric hysteresis loops (P-E) for PZT/LSMO heterostructures; the inset is remanent polarization ($P_r$) as a function of temperature. All measurements were carried out under a maximum applied field of 350 kV/cm. The $P_r$ were 54 $\mu C/cm^2$ and 32 $\mu C/cm^2$ with corresponding coercive fields of 41 kV/cm and 103 kV/cm for the heterostructures deposited on $LaAlO_3$ substrates with (HS1) LSMO $T_D=700^0C$ and (HS2) $T_D=600^0C$ respectively. In the last case (HS2) the P-E loop exhibited a slight shift toward positive voltages as a consequence of a high in built field; this is presumably due to the extra stress across the interface which in turn is probably caused by the higher annealing temperature. The same phenomenon was reported by Wu *et al.*[14] in epitaxial PZT thin films.

The magnetic (M-H) hysteresis loops of PZT/LSMO heterostructures at 300 K are shown in Figure 4, where the maximum magnetic field applied was 19 kOe. Ferromagnetic behavior is exhibited in all samples due to the LSMO layer. The average saturated magnetization value $M_s$ was ~ 208 emu/cm$^3$ for HS1 and ~ 173 emu/cm$^3$ for HS2 respectively, evidencing the close relation between the deposition temperature and ferromagnetic properties in our samples.

In Summary, artificially engineered epitaxial ferroelectric PZT and ferromagnetic LSMO hetrostructures were grown onto (100)-LAO substrates by pulsed laser deposition. Structural, electrical, and magnetic properties in these heterostructures are reported over a wide range of temperatures and frequencies. The x-ray diffraction measurements and atomic force microscopy images showed oriented and homogeneous grains with very small surface roughness (<1.8 nm). Room-temperature multiferroic properties were exhibited in the samples while the individual layers retained their ferroelectric and ferromagtnetic properties. However these properties are strongly influenced by the deposition temperature of LSMO. Frequency and temperature dependent dielectric dispersions were observed above the LSMO magnetic Curie temperature and explained in the light of extrinsic contributions and the metal-insulator transition.

**ACKNOWLEDGMENTS**

This work was partially supported by W911NF-06-1-0183 and DoE FG 02-08ER46526 grants. One of the authors (S. D.) acknowledges an IFN fellowship.

# LIST OF FIGURES

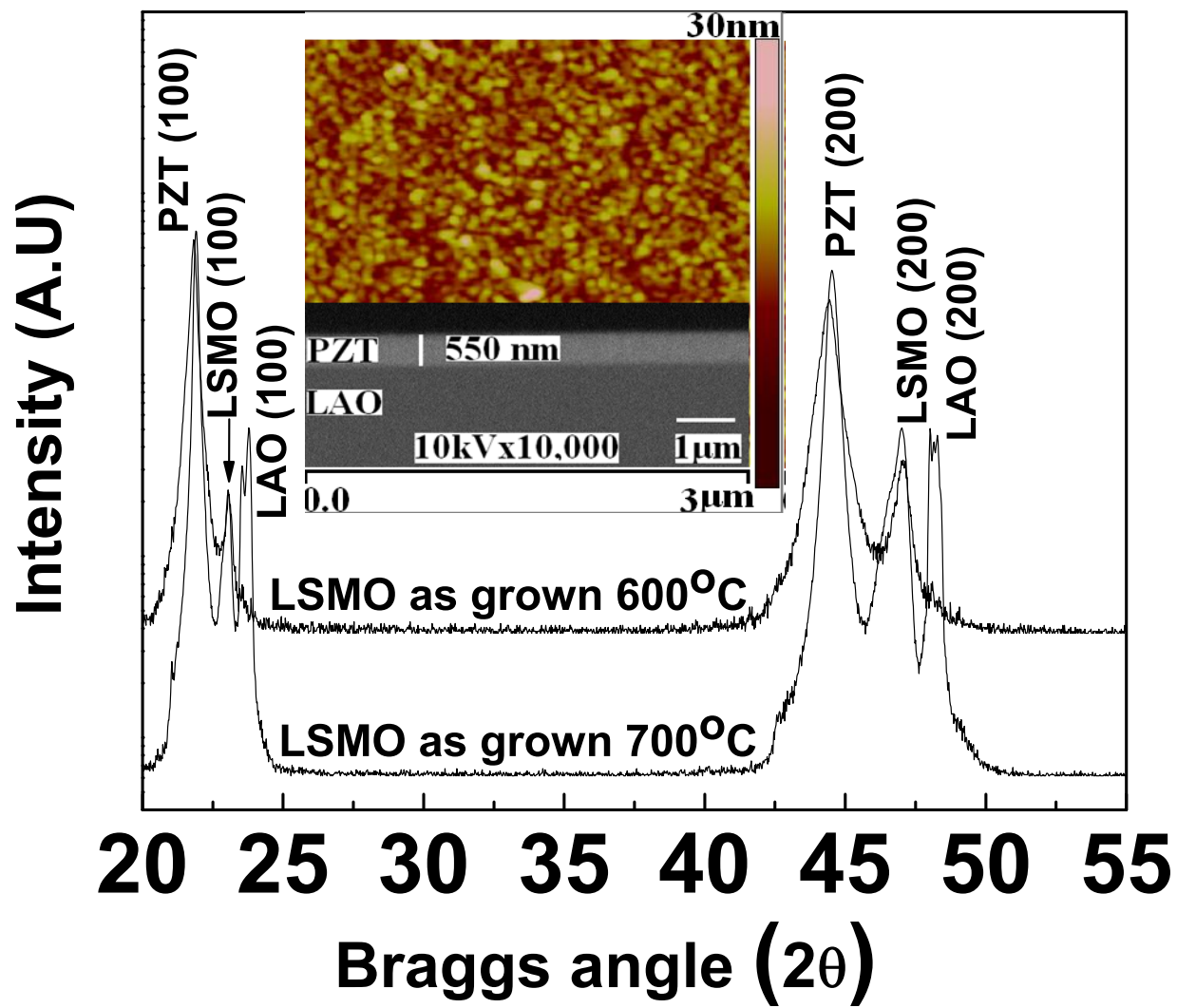

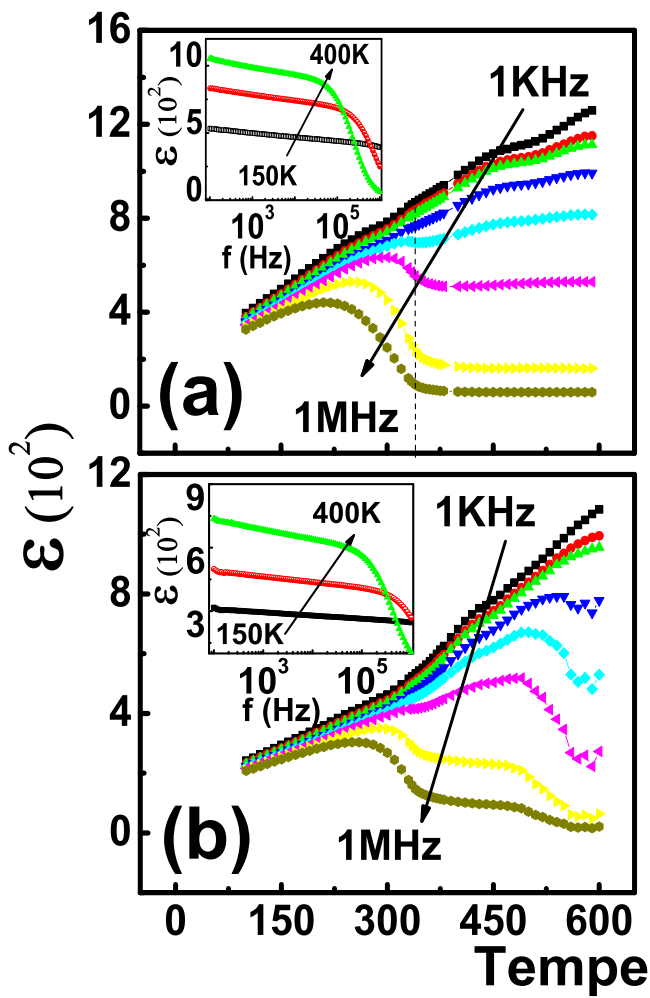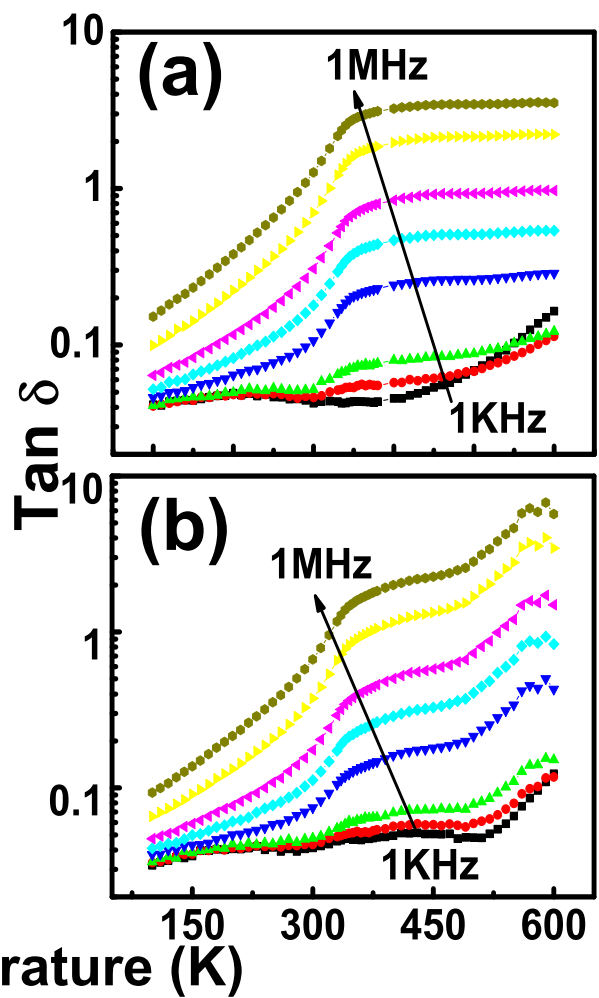

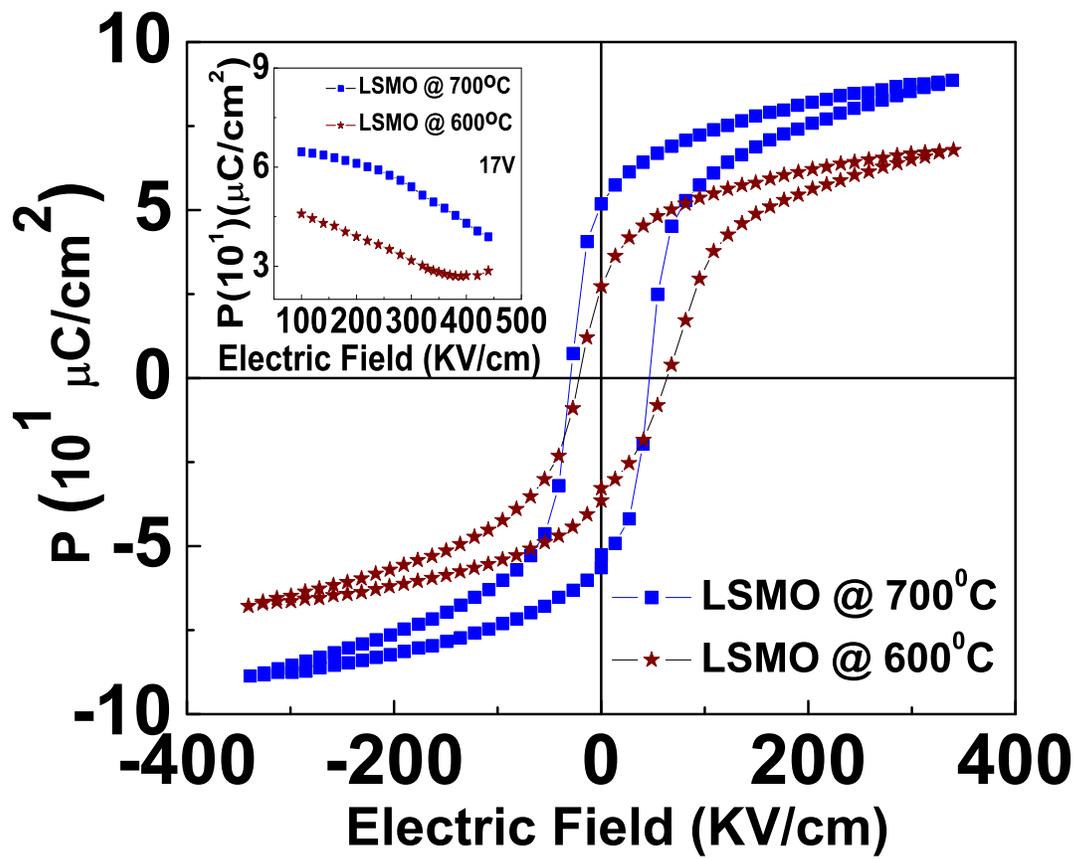

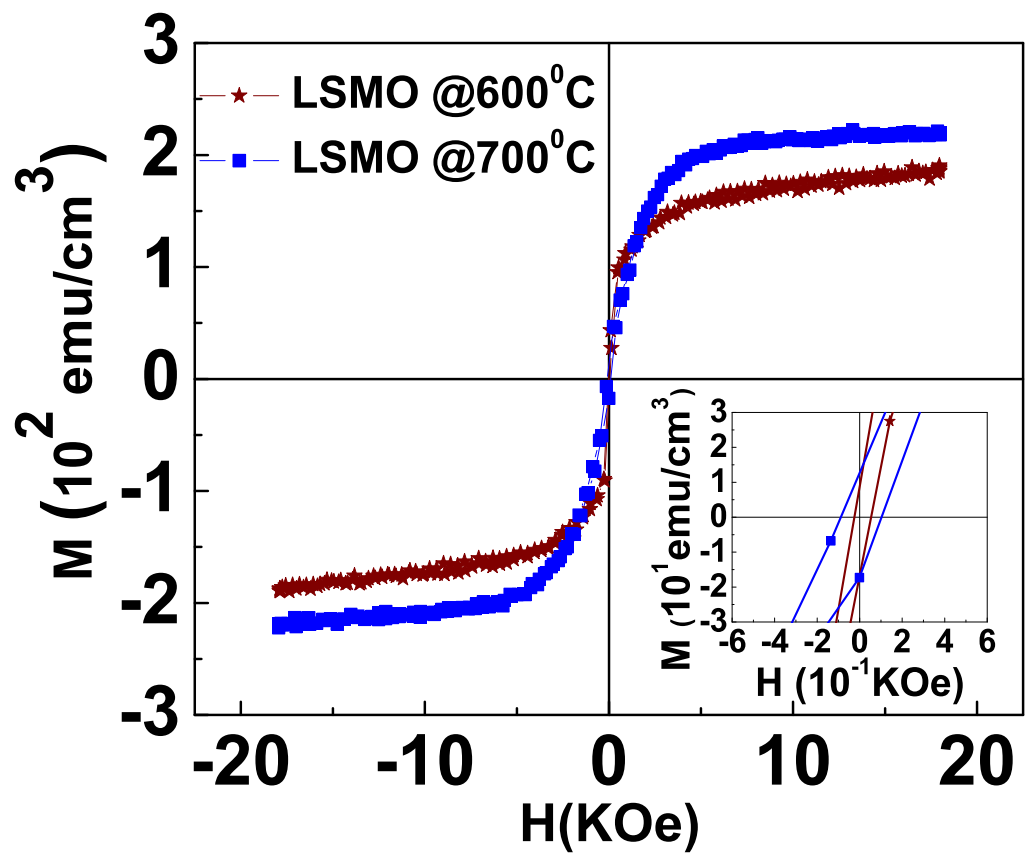